\documentstyle[12pt,bezier]{article}
\begin{document}
\begin{titlepage}
\pagestyle{empty}
\baselineskip=21pt
\rightline{UMN-TH-1201/93}
\rightline{TPI-MINN-93/16-T}
\rightline{hep-ph/yymmddd}
\rightline{April 1993}
\vskip .2in
\begin{center}
{\large{\bf On the Erasure and Regeneration \\ of the Primordial Baryon
Asymmetry by Sphalerons}} \end{center}
\vskip .1in
\begin{center}
James M.~Cline

Kimmo Kainulainen

and

Keith A. Olive

{\it School of Physics and Astronomy, University of Minnesota}

{\it Minneapolis, MN 55455, USA}

\vskip .2in

\end{center}
\vskip 1in
\centerline{ {\bf Abstract} }
\baselineskip=18pt
We show that a cosmological baryon asymmetry generated at the GUT scale,
which would be destroyed at lower temperatures by sphalerons and possible new
$B$- or $L$-violating effects, can naturally be preserved by an asymmetry in
the number of  right-handed electrons.  This results in a significant softening
of previously  derived baryogenesis-based constraints on the strength of exotic
$B$- or $L$-violating interactions.
\noindent
\end{titlepage}
\baselineskip=18pt

{\newcommand{\gtwid}{\mathrel{\raise.3ex\hbox{$>$\kern-.75em\lower1ex
\hbox{$\sim$}}}}
{\newcommand{\ltwid}{\mathrel{\raise.3ex\hbox{$<$\kern-.75em\lower1ex
\hbox{$\sim$}}}}

It was first noticed by Fukugita and Yanagida (FY) \cite{fy} that, if the
baryons comprising us were created early enough in cosmological history, then
lepton number  must be at least a fairly good approximate symmetry. They
reasoned that sphalerons, which tend to destroy any net excess in $B+L$ number
\cite{sph}, would ensure the baryon and lepton asymmetries were equal and
opposite, roughly speaking.  But if lepton violating processes were
significant, then $B$ and $L$ would both be driven to zero.  Their analysis
yielded a rather weak bound of 50 keV on the majorana mass of a neutrino.
Since then a great deal of effort has gone into strengthening and generalizing
their result. References \cite{ht} and \cite{nb} significantly improved the FY
bound by requiring the $L$-violating interactions responsible for seesaw
neutrino masses to be out of equilibrium at very high temperatures where the
sphalerons go out of equilibrium, rather than the much lower electroweak phase
transition temperature used by FY. The improved results were then applied to
general operators which violate either $B$ or $L$ \cite{cdeo12,sonia}. Although
the strongest bounds on non-renormalizable interactions of this type come from
considering their effects at the sphaleron decoupling temperature, $T_m \sim
10^{12}$ GeV, in the supersymmetric case it was realized \cite{iq} that above a
certain scale associated with supersymmetry breaking, $T_s \sim 10^8$ GeV, the
presence of new anomalies would cause the baryon number to be encoded in
supersymmetric particles, saving it from erasure until temperatures below
$T_s$. Moreover, in generic models of inflation, equilibrium is restored at the
relatively low temperature of $T_r \sim 10^5$ GeV and further softens this
bound \cite{cdo2}.

In this letter we point out a surprise: most of the efforts to strengthen the
original Fukugita-Yanagida bound are invalidated by a rather mundane feature of
the Standard Model, namely the smallness of the Yukawa coupling of the
right-handed electron, which causes interactions that change the net number of
$e_R$'s to come into equilibrium only at very late times. The authors of
ref.~\cite{cdeo3} investigated whether the equilibration was late enough so
that a baryon excess in a universe with $B-L=0$ could be preserved by an $e_R$
asymmetry, despite sphaleron effects.  Whereas their answer was ``no,'' here we
make the startling observation that the answer becomes ``yes'' if one {\it
includes} some explicit $L$ or $B$ violation, such as Majorana neutrino masses.
This is contrary to the usual prejudice toward the strength of lepton
violation vis-a-vis baryogenesis, where it was always assumed that smaller is
better.  If on the other hand $B-L$ is nonzero, lepton number violation is not
necessary, but it is much more permissible than was previously thought.

We emphasize that it is the right-handed electrons that play the role of
protector of the baryon asymmetry.  The key observation is that any asymmetry
in their own numbers remains untouched until relatively low temperatures around
$T_*=10$ TeV, when the small Yukawa interactions with left-handed electrons and
Higgs bosons finally become fast enough to convert the $e_R$'s into $e_L$'s.
Because sphalerons interact only with the left-handed particles, they can only
directly deplete the latter.  Therefore, as long as any additional lepton
violating interactions have gone out of thermal equilibrium before the
right-handed electrons come {\it into} equilibrium, the initial $e_R$ asymmetry
is protected from being washed out. When the temperature eventually falls below
$\sim$ 10 TeV, sphalerons will be able to convert a sizable fraction of the
initial $e_R$ asymmetry into the baryon excess that exists today.
The condition that lepton (or baryon) violating effects freeze out before
10 TeV leads to new, relaxed bounds on the sizes of various symmetry-breaking
operators.

We will for the most part concentrate on the $\Delta L = 2$, $D=5$
operator considered by FY. The major result is that the FY's original
50 keV bound on the neutrino mass matrix elements can only be strengthened
to a few keV.  The difficulty of arranging masses and
mixing angles to conflict with this bound will be explored, leading
to the strong conclusion that sphalerons, together with the $L$-violating
effects of the dimension five operator $LLHH$, generically cannot destroy any
primordial baryon asymmetry, which might be produced in grand unified
theories!  The bounds on other $B$ and $L$ violating operators, though still
interesting, are also weakened relative to
previous expectations.
In the remainder of our letter we elaborate upon and prove these statements. To
avoid burying the simplicity of our idea in an avalanche of details, we give a
rough derivation, whose result will be corroborated more rigorously in a future
publication \cite{cko3}.

Let us consider first the effective operator ${\lambda^2 \over 2M} LLHH$
generated by adding right-handed neutrinos $N_k$ to the  standard model
Lagrangian, with large Majorana masses $M_k$ and Yukawa couplings to the
Higgs and lepton doublets $H$ and $L_j$,
\begin{equation}
\frac{1}{2} {\bar N}_k (i{\raise.15ex\hbox{/}\kern-.57em\partial} - M_k) N_k -
\frac{1}{2}\lambda_{jk} \bar L_j H(1-\gamma_5) N_k + {\rm\ h.c.}
\label{S1}
\end{equation}
Below the electroweak phase transition, the SU(2) doublet neutrinos acquire
Majorana masses $m_{ij} = \sum_{k} \lambda_{ik} \lambda^\dagger_{kj} v^2/M_k$
and $v$ is the Higgs vacuum expectation value of 174 GeV.  Therefore the
dimension five operator induced by $N_k$ exchange may be written
\begin{equation}
\frac{1}{2v^2}\sum_{ij}m_{ij}(\bar L_iH)(H^TL^c_j).
\label{effective}
\end{equation}

Figure 1 illustrates the sequence of events that would lead to the destruction
of the baryon asymmetry, and suggests how to avoid that outcome. At very high
temperatures, lepton-violating ($\Delta L$) annihilation processes due to the
interaction (\ref{effective}) are in equilibrium because their rates scale as
$T^3$, whereas the Hubble expansion rate goes like $T^2$.  Slowest of all are
the processes shown in fig.~2 that equilibrate the left- and right-handed
electrons, scaling like $T$. These come {\it into} equilibrium at a temperature
$T_*$.  If $T_*$ is much greater than $T_f$, the freezeout temperature of the
$\Delta L$ processes, then baryon number is completely washed out by the
sphalerons.  But in the opposite case, $T_f > T_*$, sphalerons are able to
destroy baryon number only until $T_f$.  At this point, they begin to {\it
regenerate} baryon number out of the $e_L$ asymmetry, which in turn came from
the primordial $e_R$ asymmetry, once the left-right equilibrium was established
at $T=T_*$.

We are interested in the rate $\Gamma_{\Delta e_L}$ at which $e_L$-type lepton
number is violated, since electrons have the smallest Yukawa coupling and hence
the lowest temperature for reaching right-left equilibrium:
\begin{equation}
\Gamma_{\Delta e_L} = {11\over 32\pi^3}{\mu^2 T^3\over v^4}; \qquad
\mu^2\equiv 2|m_{ee}|^2+|m_{e\mu}|^2+|m_{e\tau}|^2.
\label{gammaL}
\end{equation}
This is just the thermally averaged rate of $\Delta L = 2$ annihilations $L_e
L_i\to HH$, summed over families $i$.\footnote{In arriving at the coefficient,
e.g.~ for the $ee$-part, we considered the channels $e_Le_L \leftrightarrow
h^-h^-$, $e_L\nu_{eL} \leftrightarrow h^-h_0^*$, $e_Lh^+ \leftrightarrow
e_L^ch^-$ and $e_Lh^+ \leftrightarrow h_0^*\nu_{eL}^c$. We used initial state
Maxwell-Boltzmann distribution functions but ignored those of the final states.
The factor of 2 difference between the $ee$ and $e\mu$ or $e\tau$ contributions
is due to the fact that the $ee$ processes change $e_L$ lepton number by two
units, twice as much as the other processes.} Note that the mass matrix is
generally not diagonal, as we are in the flavor eigenstate basis.  Comparing
this rate to that of the Hubble expansion ($\cong 17 T^2/M_P$)
yields a freezeout temperature of
\begin{equation}
T_f =120\left({{\rm keV}^2\over\mu^2}\right){\rm\ TeV}.
\label{freezeout} \end{equation}

On the other hand, the rate of $e_R$-$e_L$ equilibration is determined by the
Higgs decays and inverse decays, shown in fig.~2.  There will also be
scattering processes such as $t_R\bar t_L\to e_R\bar e_L$, but we have found
that for Higgs boson masses above the experimental limit they are subdominant
to the decays, whose rate is \cite{cdeo12}
\begin{equation}
\Gamma_d = {\pi h_e^2\over 192\zeta(3)}{m_h^2(T)\over T},
\label{decayrate}
\end{equation}
where $h_e = 2.9\times 10^{-6}$ is the electron Yukawa coupling. The thermal
Higgs mass can be parametrized in the form $m_h(T) \cong xT$, with $x=
m_0/2T_{EW}$, in terms of the vacuum Higgs mass $m_0$ and the critical
temperature of the electroweak phase transition $T_{EW}$. Laboratory bounds on
$m_0$ thus imply that $x>0.4$.  The thermal masses of the electrons are much
smaller and have been neglected.  Again comparing to the expansion rate, we
find that the decays come into equilibrium at a temperature
\begin{equation}
T_* = 80 x^2 {\rm\ TeV}.
\label{Tstar}
\end{equation}
This temperature scale is significantly  below the sphaleron scale $T_m$ and
even the supersymmetric scale $T_s$. Though it appears to be only a modest
improvement over the inflationary scale $T_r$, our determination of $T_*$
carries none of the uncertainties necessarily carried by the inflationary
bound.

Previous work took as the criterion for preserving the baryon asymmetry that
$T_f$ must exceed some very high temperature, where the baryons were first
produced or the sphalerons first came into equilibrium. Our criterion is much
weaker; we demand only that $T_f > T_*$, assuming that the asymmetry of $e_R$'s
is of the same order as the original baryon asymmetry, which generally happens
in GUT baryogenesis scenarios.   This condition on the freezeout temperatures
translates into a bound on the neutrino mass matrix elements,
\begin{equation}
\mu \equiv \left(2|m_{ee}|^2 + |m_{e\mu}|^2 + |m_{e\tau}|^2\right)^{1/2} <
1.2\; x^{-1} {\rm\ keV}.
\label{newbound}
\end{equation}

Because we have allowed for mixing between the neutrino flavors, we do not
obtain a direct bound on one of the neutrino masses as in ref.~\cite{fy}. To
rephrase (\ref{newbound}) in terms of potentially measurable quantities, first
note that $m_{ee}$ is precisely the quantity constrained to be less than 1 eV
by searches for neutrinoless double beta decay \cite{bay}, and so it is
negligible in (\ref{newbound}).  It is also convenient to assume that the
mixing angles $\theta_{i\alpha}$ and $CP$-violating phases that arise from the
diagonalization of $m_{ij}$ are small, and that there is a hierarchy of mass
eigenvalues. (If any of the mixing angles was large, the corresponding mass
eigenvalue would have to be so small as to be irrelevant in equation (7).)
Then $\mu$ is simply related to $\theta_{e\alpha}$ and the masses $m_\alpha$ by
\begin{equation}
\mu^2 \cong \theta^2_{e2}m_2^2 + \theta^2_{e3}m_3^2,
\label{musquared}
\end{equation}
plus corrections of order $\theta_{e\alpha}^4$.  One quickly realizes that
the bound (\ref{newbound}) guaranteeing preservation of the baryon asymmetry
must already be satisfied due to other constraints, unless the standard model
is supplemented with some way for heavy neutrinos to decay much faster.  If
$m_2$ or $m_3$ is in the keV region, the corresponding neutrino will make
too large a contribution to the energy density of the universe, since only
the radiative decay mode $\nu\to\gamma\nu_1$ is available, and this is far too
slow.  If $m_3>2m_e$ then the weak decay $\nu_3\to e^+e^-\nu_1$ can occur, with
a lifetime of
\begin{equation}
\tau_3 \cong 9\times 10^{-4}\left({31{\rm\ MeV}\over m_3}\right)^5
\theta_{e3}^{-2} {\rm\ s}.
\label{lifetime}
\end{equation}
The experimental limit on this mixing angle is $\theta_{e3}^2<10^{-6}$ for a 31
MeV neutrino, and it relaxes to $\theta_{e3}^2<10^{-4}$ if $m_3 = 3.5$ MeV
\cite{britton}. The respective lifetimes are therefore 900 s and $5\times 10^5$
s.  Although these are compatible with the expansion of the universe, they are
in strong conflict with nucleosynthesis \cite{kolb} and cosmic microwave
background constraints.

A remarkable feature of our result, which provides a new means for generating
the baryon asymmetry, is that it is {\it independent} of the initial values of
$B$ and $L$, and it applies so long as three requirements are satisfied: (1)
the existence of a primordial $e_R$ asymmetry comparable to the present baryon
asymmetry, (2) the absence of any exotic interactions involving $e_R$ which
would wash out this asymmetry before the electroweak phase transition, and (3)
additional $B$- or $L$-violating interactions which must have frozen out
between $10^{12}$ GeV, when sphalerons reached equilibrium, and
$T_*$, the temperature at which $e_R$-$e_L$ equilibrium began.\footnote{To be
precise, this statement applies specifically to baryon or
{\it electron}-type lepton number. Violation of other varieties $L$ can
remain in equilibrium even below $T_*$, resulting in a different, yet nonzero
final baryon asymmetry, as will be shown in \cite{cko3}.}\  Consider again the
case of the dimension five $\Delta L=2$ operator.  Under the above conditions,
the $L$-violating and sphaleron reactions will establish equilibrium between
all interacting species, with the boundary condition that the $e_R$ asymmetry
is conserved. Because $e_R$ carries charge, this constraint carries over to the
interacting species because the universe is charge-neutral. One can easily show
that
\begin{eqnarray} B &=& 12\mu_{q_L};\nonumber \\ L &=&
3(\mu_{\mu_L} + \mu_{\tau_L} ) + 2\mu_{e_L} + \mu_{e_R} - 2\mu_h;\\ Q &=&
6\mu_{q_L} - 2(\mu_{\mu_L} + \mu_{\tau_L} )  - \mu_{e_L} - \mu_{e_R} +
13\mu_h,\nonumber \label{chempot}
\end{eqnarray}
where $\mu_{q_L}$ and $\mu_h$ are the chemical potentials for left handed
quarks and Higgs bosons and Q is the total charge. Imposing $Q = 0$ along with
constraints coming from the sphaleron processes, $9\mu_{q_L} + \sum_{i=1}^3
\mu_i = 0$, and lepton number violation, $\mu_{i} = - \mu_h$ ($i \equiv e_L,
\mu_L, \tau_L$), one finds that just above $T_*$, $B$ and $L$ are given in
terms of the primordial $e_R$ asymmetry $\mu_{e_R,p}$
\begin{equation}
B_* = \frac{1}{5}\mu_{e_R,p}; \qquad L_* = \frac{1}{2}\mu_{e_R,p},
\label{BandL}
\end{equation}
{\em regardless} of the primordial values of $B$ and $L$. Assuming there was no
lepton number violation below $T_*$ would then lead to a final baryon number
completely determined by $\mu_{e_R,p}$,
\begin {equation}
B_f = \frac{28}{79}(B_*-L_*) \cong -0.11 \mu_{e_R,p},
\label{Bfinal}
\end{equation}
Note that for the particular case that $B-L$ was initially zero, lepton (or
baryon) number violation at some intermediate scale is actually {\it helpful}
for generating a baryon asymmetry, rather than being constrained. The strength
of $L$-violation needed is quite modest: all we require is that either an
$L$-violating decay or scattering remain in equilibrium to temperatures below
$T_m \sim 10^{12}$ GeV.  This could be accomplished by a tau neutrino mass of
at least $10^{-2}$ eV \cite{cko1} in the case of decays, or a slightly larger
mass of 0.1 eV in the case that $\Delta L=2$ scatterings dominate over decays
\cite{ht}.

Although we have focused on the $\Delta L = 2$ operator $LLHH$, the same
condition must be applied to other $B$ or $L$ violating operators, namely
those interactions must go out of equilibrium at $T > T_*$. Indeed there is a
large number of potentially dangerous operators in grand unifed or
supersymmetric theories, bounds on which were considered in detail in
refs.~\cite{cdeo12,sonia,iq,dr}.  Our arguments above modify the results for
all dimension $D > 5$ operators. For renormalizable operators, {\it e.g.,}
those
found in supersymmetric models with $R$-parity violation, the interaction rate
increases slower than the Hubble rate as a function of temperature, and the
bound is derived using the {\it lowest} temperature at which sphalerons are in
equilibrium, $T \sim 100$ GeV. For a nonrenormalizable operator of the form
$M^{-n}O_D$ with $D=4 + n$, the strongest bound is based on the {\it
highest} temperature the two interactions ($B$ or $L$ violation, and
$e_L\leftrightarrow e_R$) are both in equilibrium
\cite{cdeo12},
\begin{equation}
M^{2n} \gtwid \frac{1}{17} M_P T_*^{2n-1}
\label{newoldbound}
\end{equation}
Because of the smallness of the electron Yukawa coupling, we must take $T_*$
as the maximum temperature, giving a weaker condition than previous derivations
that overlooked this fact.

An interesting application of this limit is to a $\Delta B = 2$, $D=9$
operator that would induce neutron-antineutron oscillations \cite{cdeo12}.
Using $T_* \sim 10$ TeV from (\ref{Tstar}) we find that the limit on the heavy
mass scale $M$ suppressing the operator becomes $M \gtwid 3 \times 10^5$ GeV,
which is substantially weaker than the stronger of the two bounds
quoted in \cite{cdeo12} and is
comparable to the current experimental bound of $M > 10^5 - 10^6$ GeV.

Our mechanism for the generation of a baryon asymmetry also applies to general
$B$ or $L$ violating operators.  For example the operator
$M^{-5}(u_Rd_Rd_R)^2$, which would induce $n$-${\bar n}$ oscillations, imposes
the condition $\mu_{u_R} + 2\mu_{d_R} = 0$ on the chemical potentials while it
is in equilibrium.  Supposing this was true only for $T>T_*$, so that the limit
(\ref{newoldbound}) was satisfied, one would find the final baryon number to be
proportional to the difference of the primordial $e_L$ and $e_R$
asymmetries,\footnote{Because of the additional residual conserved quantities
$L_e - L_\mu$ and $L_e - L_\tau$, this difference is also conserved.}\ $B_f =
\frac{8}{79}(\mu_{e_L,p}-\mu_{e_R,p})$. Similar results would be obtained for
other operators. It could also happen that interactions of more than one
operator were in equilibrium simultaneuosly, but since the interpretation of
such situations is somewhat more convoluted, we will postpone their discussion
until the forthcoming publication \cite{cko3}.

We conclude by stressing that our mechanism for regenerating the baryon
asymmetry from an $e_R$ asymmetry is generic, in the sense that it is natural
for decaying GUT gauge bosons or supersymmetric condensates to produce as large
an excess of $e_R$'s as of any other particle.  Although there are many
possible effective interactions that could destroy the $e_R$ asymmetry, we have
noted that the most popular one is already so constrained by laboratory
experiments that it cannot interfere with our scenario.  This is the dimension
five operator (\ref{effective}) that would give seesaw masses to neutrinos.
More generally, we have shown that other interactions threatening the baryon
asymmetry are impotent unless they stay in equilibrium below temperatures
around 10 TeV, the temperature at which an $e_R$ excess becomes vulnerable to
being erased, so that these interactions are less tightly constrained than was
formerly supposed.
\vskip 1.0truecm
\noindent {\bf Acknowledgements}
\vskip 1.0truecm
We would like to thank Larry McLerran and Sonia Paban for useful discussions.
This work was supported in part by  DOE grant DE-AC02-83ER-40105.
The work of KAO was in addition supported by a Presidential Young
Investigator Award.

\newpage
}}

\unitlength=1.00mm
\thicklines
\begin{picture}(131.43,133.69)
\put(109.67,100.00){\line(1,0){1.92}}
\put(112.98,100.12){\line(1,0){1.76}}
\put(116.11,100.12){\line(1,0){1.88}}
\put(119.25,100.12){\line(1,0){1.88}}
\put(121.13,100.12){\line(5,6){8.28}}
\put(121.13,100.12){\line(5,-6){8.41}}
\put(108.21,100.12){\line(-1,0){1.76}}
\put(105.07,100.12){\makebox(0,0)[cc]{H}}
\put(131.43,112.16){\makebox(0,0)[cc]{$e_R$}}
\put(131.43,87.76){\makebox(0,0)[cc]{$\bar L_e$}}
\put(104.67,76.79){\makebox(0,0)[lc]{Fig.~2. Decay of Higgs boson}}
\put(104.67,72.00){\makebox(0,0)[lc]{into $e_R$ and $L_e$ doublet.}}
\put(40.00,89.99){\line(0,1){0.00}}
\put(10.00,90.00){\line(0,1){39.99}}
\bezier{208}(17.77,130.03)(30.00,100.00)(50.00,100.00)
\bezier{172}(30.00,130.00)(40.00,100.00)(48.73,92.88)
\bezier{160}(11.80,121.51)(30.00,110.00)(48.86,110.54)
\put(10.03,133.42){\makebox(0,0)[cc]{rate}}
\put(67.74,90.18){\makebox(0,0)[cc]{$T^{-1}$}}
\put(51.27,99.88){\makebox(0,0)[lc]{Hubble}}
\put(51.27,110.24){\makebox(0,0)[lc]{left-right}}
\put(28.68,133.69){\makebox(0,0)[cc]{$\Delta L$}}
\put(26.27,91.66){\line(0,-1){3.05}}
\put(26.27,84.66){\makebox(0,0)[cc]{$T_*^{-1}$}}
\put(52.11,91.66){\line(0,-1){3.05}}
\put(52.11,84.96){\makebox(0,0)[cc]{$T_{EW}^{-1}$}}
\put(41.99,91.79){\line(0,-1){3.57}}
\put(42.05,84.66){\makebox(0,0)[cc]{$T_f^{-1}$}}
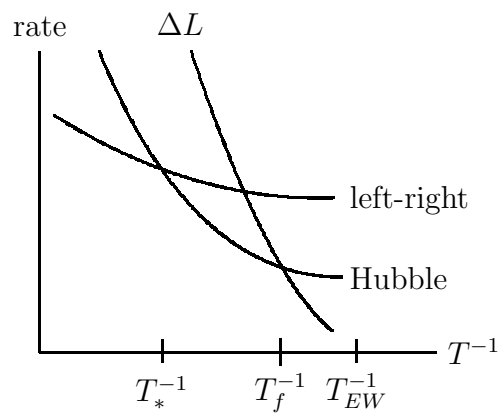
\put(10.00,76.79){\makebox(0,0)[lc]{Fig.~1. Lepton-violating and left-right}}
\put(10.00,72.00){\makebox(0,0)[lc]{equilibrating rates.}}
\put(10.00,90.00){\line(1,0){52.77}}
\end{picture}

\end{document}